\documentclass[12pt,tightenlines,eqsecnum,floats,aps,amsmath,amssymb,tightenlines,nofootinbib,prd,shownopacs,notitlepage,superscriptaddress]{revtex4-1}

\usepackage{graphicx,verbatim}
\usepackage{amsmath}
\usepackage{amsfonts}
\usepackage{amssymb}
\usepackage[colorinlistoftodos]{todonotes}
\usepackage{epstopdf}
\usepackage{psfrag}
\usepackage{accents}
\usepackage{mathrsfs}
\usepackage{bm}
\usepackage[applemac]{inputenc}
\usepackage[spanish,english]{babel}
\usepackage{amsmath,amssymb,amsfonts,cancel}
\usepackage[colorinlistoftodos]{todonotes}
\usepackage{epstopdf}

\usepackage{color}
\usepackage{soul}
\usepackage{hyperref}
\usepackage{slashed}

\allowdisplaybreaks

\newcommand{\be}{\begin{equation}}
\newcommand{\ee}{\end{equation}}
\newcommand{\bea}{\begin{eqnarray}}
\newcommand{\eea}{\end{eqnarray}}

\newcommand{\pb}[1]{\hbox{\lower0.5ex\hbox{${}_{\leftarrow}$}}\kern-1.9ex{#1}}

\def\be{\begin{equation}}
\def\ee{\end{equation}}
\def\ba{\begin{eqnarray}}
\def\ea{\end{eqnarray}}

\def\h{\hat}
\def\b{\bar}

\def\f{\frac}

\def\rmd{\mathrm{d}}

\def\go{g^{\circ}}
\def\Boxo{\mathring{\Box}}
\def\Ho{H_{\circ}}
\def\phio{\mathring{\phi}}
\def\bfphi{\mathring{\boldsymbol{\varphi}}}
\def\omegao{\mathring{\omega}}
\def\omegap{\omega_{\rm Pl}}

\def\ha{\hat{A} (\omega)}
\def\had{\hat{A}^\dag (\omega)}
\def\bcoh{\langle\Psi_{\phio}|}
\def\kcoh{|\Psi_{\phio} \rangle}
\def\d{\dagger}
\def\e{\exp\, }
\def\F{\mathcal{F}}
\def\H{\mathcal{H}}
\def\N{{\cal N}} 

\def\o{{\omega}}

\begin{document}
\title{Exploring quantum geometry created by quantum matter}
\author{Abhay Ashtekar}\email{ashtekar.gravity@gmail.com}
\affiliation{Physics Department \& Institute for Gravitation and the Cosmos\\ The Pennsylvania State University, University Park, PA 16802 U.S.A}

\begin{abstract}

Exactly soluble models can serve as excellent tools to explore conceptual issues in non-perturbative quantum gravity. In perturbative approaches, it is only the two radiative modes of the linearized gravitational field that are quantized. The goal of this investigation is to probe the `Coulombic' aspects of quantum geometry that are governed entirely by matter sources. Since there are no gravitational waves in 3 dimensions, 3-d gravity coupled to matter provides an ideal arena for this task. Our analysis will reveal novel aspects of quantum gravity that bring out limitations of classical and semi-classical theories in unforeseen regimes: non-linearities of general relativity can magnify small quantum fluctuations in the matter sector to large effects in the gravitational sector. Finally, this analysis leads to thought experiments that bring out rather starkly why understanding of the nature of physical reality depends sensitively on the theoretical lens with which it is probed. As theories becomes richer, new scales emerge, triggering novel effects that could not be imagined before. The model provides a concise realization of this well-known chain.

\end{abstract}

\maketitle

\section{Introduction}
\label{s1}

This work has several different motivations, stemming from various aspects of quantum gravity. A number of bold ideas have been put forward in 4-d quantum gravity that appear to be plausible from one perspective but puzzling --and even unsettling-- from another. Exactly soluble models with local degrees of freedom --sometimes called midi-superspace in the literature-- can be used to examine these issues \emph{non-perturbatively}. Of course, one has to make simplifying assumptions to arrive at these models. Therefore, can miss important aspects of the full theory and one has to exercise due caution in assessing the conclusions one arrives at. Nonetheless, these models can provide pointers and bring to forefront new conceptual issues in a crisp manner.

One such issue concerns the potential implications of `trans-Planckian modes' associated with extremely high frequency fields that arise, e.g., in the analysis of Hawking radiation. Over the years, there have been suggestions that an adequate handling of these modes may lead to novel physics by, e.g., having to abandon local Lorentz invariance (see, e.g., \cite{sctj}). Exactly soluble models provide a natural arena to analyze this issue. Models we will discuss, for example, do have unforeseen effects induced by high frequency modes. It is then natural to ask if they lead to a loss of local Lorentz invariance in the final non-perturbative theory, or if this symmetry is preserved by the new physics. Another example comes from the widely held view that the usual counting of states in quantum field theory becomes inadequate at high frequencies and the correct counting would lead to a holographic picture in which physics in the bulk is fully captured by states residing on the boundary (see, e.g., \cite{jb,ls,gth} for early works and \cite{hol-review} for a recent review). Is this view supported in exactly soluble models?  

A third example comes from another commonly held view --first introduced by Wheeler \cite{jaw}-- that there is a `space-time foam' at the microscopic level because of the perpetual quantum fluctuations of the gravitational field.  In Wheeler's paradigm, these fluctuations are to emerge as novel non-perturbative effects associated with the quantum nature of geometry, rather than from the ultraviolet behavior of graviton-mediated interactions in perturbative approaches. This view then leads one to visualize the `non-perturbative ground state' of full quantum gravity as having very different micro-structure than that suggested by the smooth, tame geometry of Minkowski space and the associated perturbative vacuum. Is this idea borne out in exactly soluble models? If so, the standard positive energy theorem of classical general relativity may not extend to non-perturbative quantum gravity; the energy density in the ground starts could well be Planck density with either sign. Is this what happens? 

Exactly soluble models can also be well-suited to probe an issue at the opposite corner of the theory: semi-classical gravity. This is a mathematically well-defined theory in which gravity/geometry is treated classically, matter fields quantum mechanically, and the two are related by semi-classical Einstein's equations in which the right hand side is provided by the expectation value of stress-energy of quantum matter. It is generally assumed that the theory would provide a good approximation to the predictions of quantum gravity so long as space-time curvature is low compared to the Planck scale.  Exactly soluble models can shed light on the validity of this conjecture. Can one arrive at this theory from full quantum gravity in the appropriate regime? Are there potential surprises about its domain of validity?

Another motivation for this work comes from the ongoing debate on whether quantum matter can inject genuinely quantum information into the gravitational field, or, if it is only the radiative --or `true'-- degrees of the gravitational field that are quantum mechanical (see, e.g. \cite{boseetal,cmvv,cbp,dgw}). The soluble models we consider can be thought of as 3-dimensional general relativity coupled to the Maxwell (or Klein Gordon) fields. The 3-dimensional viewpoint provides a conceptually clean arena to probe this issue since the gravitational field is now completely determined by matter; there are no (gravitational waves or) gravitons in 3 dimensions. Nonetheless, we will find that quantum geometry does exhibit specific and quintessentially quantum features. In fact the nonlinear coupling between matter and gravity can magnify small quantum fluctuations in the matter sector to large quantum effects in space-time geometry, leading to consequences that could not have been foreseen classically.

Finally, these soluble models also suggest  thought experiment that yield more general conceptual insights. Our understanding of physical reality is deeply influenced by the theoretical paradigm we choose to work in. Indeed, possibilities we envisage and experiments we design to test subsequent hypotheses are dictated this paradigm. For the specific models we will discuss, two constants of Nature play a fundamental role: Planck's constant $\hbar$ that dictates the size of quantum effects and Newton's constant $G$ that governs gravitational phenomena. We will find that `switching them on and off' results in drastic changes in our understanding of the underlying physical reality. 

The material is organized as follows. In section \ref{s2} we discuss the classical `midi-superspace models' and in section \ref{s3} the unforeseen quantum gravity effects they present us with. Section \ref{s4} summarizes the main results and puts them in a broader context. The key mathematical results presented here were summarized in a Letter \cite{aa-prl} quite some time ago. This article provides detailed derivations --that have been been circulated only in private communications so far-- and examines them from different angles, putting them in a broader conceptual perspective. Since significant time has lapsed, I will also discuss related results in other works that have appeared since (in particular in \cite{schmidt,rgjp,mena,beetle,admt,boseetal,cmvv,cbp,dgw,abs1,abs2}), as well as the thought experiment, mentioned above, that has not appeared in the literature.

In the main discussion we will work with 2+1 dimensional general relativity coupled with a scalar or a Maxwell field. Space-time metric $g_{ab}$ has signature -,+,+ and its derivative operator is denoted by $\nabla$; its Riemann tensor is defined by $R_{abc}{}^d k_d = 2 \nabla_{[a} \nabla_{b]} k_c$,\, the Ricci tensor by $R_{ac} = R_{abc}{}^b$,\, its scalar curvature by $R = g^{ac}\, R_{ac}$, and its Einstein tensor by $G_{ab}$. We will set $c=1$ but display the dependence of various quantities on $G$ and $\hbar$. For convenience of the reader we note that, in 3-d, Newton's constant has dimensions\,\, $[G]\,\sim\, M^{-1}$.\, Therefore, unlike in 4-d general relativity, there is a natural mass scale in the classical theory. But one cannot associate a given mass $M_{\circ}$ with a length; there is no notion of `Schwarzschild radius' of $M_{\circ}$. It is also helpful to keep in mind the differences that arise in quantum gravity from the more familiar 4-d situation. While there is a natural notion of Planck length\,\, ($[\hbar G] \sim L$)\,\, and therefore of Planck frequency, there is no notion of Planck mass that would be relevant to quantum gravity. In section \ref{s2} we will briefly discuss how the 3-d models we consider arise from a symmetry reduction of 4-d general relativity, a la Kaluza-Klein. There the 4-dimensional fields will explicitly carry a prefix $4$.

\section{Exactly soluble Models}
\label{s2}

In this section, we first recall the model we will primarily work with and then briefly discuss other similar models that have been considered in the literature. These are often referred to as `midi-superspace models' because they result from a symmetry reduction of 4-dimensional general relativity but have an infinite number of true degrees of freedom (in contradistinction with the `mini-superspace' of homogeneous cosmological models which have only a finite number of degrees of freedom).  While these models and their structure has been discussed in the literature, we review the salient features for completeness because the paper is intended for a broad audience. This discussion will also enable us to fix the notation used subsequently.

As mentioned above, in most of this paper we will work primarily in 3 space-time dimensions and with signature (-,+,+). Let us denote this space-time by $(M, g_{ab})$. $M$ will be assumed to be topologically $R^3$ and our metric $g_{ab}$ will be asymptotically flat both at null and spatial infinity in the sense of Refs. \cite{aamv,abs1,abs2} respectively. Matter fields will be taken to be either Maxwell fields $F_{ab}$ or Klein-Gordon fields $\Phi$. As is well-known, in 3-dimensions there is a duality between the two. Maxwell's equations imply that $F_{ab}$ satisfies Maxwell's equations if an only if its the dual ${}^\star\!F_a := \f{1}{2}\, \epsilon_a{}^{bc} F_{bc}$ is exact\, --i.e., ${}^\star\!F_a =: \nabla_a \Phi$--\, where $\Phi$ satisfies the Klein-Gordon equation $\nabla^a \nabla_b \Phi =0$. (Here we have used the fact that the topology of $M$ is trivial). Interestingly the dictionary ${}^\star\!F_a := \nabla_a \Phi$ translates the stress-energy tensor $T_{ab}^{\rm Max}$ to the stress-energy $T_{ab}^{\rm KG}$:
\ba T_{ab}^{\rm {Max}} &\equiv& F_a{}^m\, F_{bm} - \f{1}{4} g_{ab}\,F_{mn} F^{mn} =\,  {}^\star\!F_a\, {}^\star\!F_b - \f{1}{2} {}^\star\!F^m {}^\star\!F_m\, g_{ab} \nonumber\\ 
&=& \nabla_a\Phi\,\nabla_b \Phi - \f{1}{2} \nabla^m\Phi \nabla_m\Phi\, g_{ab}\, \equiv\, T_{ab}^{\rm KG}\, . \ea
Therefore, the pair $(g_{ab},\, \Phi)$ satisfies the coupled Klein-Gordon \emph{and} Einstein's equations if and only if the pair $(g_{ab},\, F_{ab})$ satisfies the Einstein-Maxwell equations. From notational simplicity, we will work with a Klein-Gordon field $\Phi$ in calculations. But conceptually it will often be more convenient to to regard the source as a Maxwell field $F_{ab}$ and its quantum excitations as photons. 

Thus, the Klein-Gordon field $\Phi$ and the space-time metric satisfy the coupled set of equations:
\be \label{fe} \Box \Phi =0 \qquad{\rm and,}\qquad   G_{ab} = 8\pi G\, T_{ab}\quad  \Leftrightarrow \quad R_{ab} = 8\pi G\,\nabla_a\nabla_b \Phi \, .\ee  
Hamiltonian analysis of this system has also been carried out in some detail \cite{mv}. But because (\ref{fe}) are rather complicated partial differential equations, we still do not have good control on detailed properties of their solutions.  However, it is well-known that equations simplify greatly if we restrict ourselves to the axisymmetric sector of this system consisting of solutions in which $g_{ab}$ admits a rotational Killing field $\partial/\partial\theta$, with regular axis, and $\Phi$ is Lie-dragged by it. Then one can cast the space-time metric in the form 
\be \label{g1} g_{ab}\,\rmd x^a\, \rmd x^b = e^{G\Gamma(R, T)} \, (\rmd T^2 + \rmd R^2)\, +\, R^2 \rmd \theta^2\,  \ee
where $T \in (-\infty, \infty)$ and $0 \le R < \infty$. The chart is unique up to the transformation $T \to T + \rm{const}$, and 
by inspection $R$ is the norm of the rotational Killing field. Note that if $\Gamma (R,T)=0$ the metric becomes the Minkowski metric $\go_{ab}$. The second consequence of the assumption of axisymmetry is technically more powerful: it is easy to check that a scalar field $\Phi$ satisfies the Klein-Gordon equation $\Box \Phi = 0$ w.r.t. the metric $g_{ab}$ if and only if it satisfies $\Boxo \Phi =0$ w.r.t. the Minkowski metric $\go_{ab}$. Therefore, the two equations in (\ref{fe}) now decouple: we can just solve the wave equation $\Boxo \Phi =0$ in Minkowski space, obtain a solution $\Phi$ and then use it in $R_{ab} = 8\pi G \, \nabla_a \Phi\nabla_b \Phi$ to determine the only unknown metric coefficient, $\Gamma(R,T)$. The final simplification is that this last equation can be readily solved to obtain
\be \label{sol1} \Gamma (R, T) = \f{1}{2} \, \int_0^R \rmd \b{R} \, \b{R}\,\,\, [\,(\partial_T \Phi)^2 + (\partial_{\b{R}} \Phi)^2\, ](\b{R},T)\, . \ee 
Note that the right side has a simple interpretation: It is simply the energy of the scalar field $\Phi$ \emph{w.r.t. the Minkowski metric} in a disc of radius $R$ at time $T$. Therefore, in the axisymmetric case, there is a very large class of asymptotically flat solutions  on which we have excellent control \cite{abs1,abs2}. \emph{This is our midi-superspace.} 

To summarize, in the midi-superspace of interest, to solve the field equations (\ref{fe}), one only needs an axisymmetric  solution $\Phi(R,T)$ to the Klein-Gordon equation in \emph{Minkowski space-time} $(M,\go_{ab})$. Given $\Phi$, we can simply write down the metric (\ref{g1}) such that the pair $(g_{ab}, \Phi)$ solves the coupled Einstein-Klein-Gordon equations. This is the \emph{general} solution in the axisymmetric sector of 3-dimensional gravity coupled to a Klein-Gordon (or Maxwell) field. Finally, as is well-known a general solution $\Phi$ of $\Boxo \Phi=0$ can be readily constructed by going to the frequency domain: It is given as an expansion in the basis $f^+_\omega (R,T) := J_0(\omega R) e^{-i\omega T}$ of positive frequency solutions, where $J_0$ is the zeroth order Bessel function of first kind:
\be \label{sol2} \Phi(R,T) = \int_0^\infty \rmd \omega\, \big[\phi(\omega) f^+_\omega (R,T)\, +\, \bar\phi (\omega) \bar{f^+}_\omega (R,T)\big]\, . \ee
This is the precise sense in which the midi-superspace under consideration is an exactly soluble sector of 3-dimensional gravity (and, as we will see, also of 4-d vacuum gravity). For each choice of a regular function $\phi(\omega)$ we obtain a solution $\Phi(R,T)$  to $\Boxo \Phi =0$ which then determines the only unknown coefficient $\Gamma (R,T)$ in the metric (\ref{g1}) such that $(\Phi, g_{ab})$ satisfy the coupled Einstein-Klein-Gordon system (\ref{fe}).  This is a midi-superspace because the system has one local degree of freedom that is neatly coded in the function $\phi(\omega)$ in the above formulation. 

To make the conceptual considerations sharper, will often focus on solutions in which $\Phi(R,T)$ has initial data of compact support on a Cauchy slice (although this restriction can be relaxed by allowing the much milder fall-off specified in \cite{aamv}). It then follows that the $\Phi(R,T)$ vanishes outside some radius $R= R_\circ(T)$, say, for each $T$. Denote by $M^\prime$ the complement of the support of the given solution $\Phi$.  In this region $M^\prime$ we have: 
\ba\Gamma (R, T) &=& \f{1}{2} \, \int_0^\infty \rmd R \, R\,\, [(\partial_T \Phi)^2 + (\partial_R \Phi)^2](R,T)\,\nonumber\\
&=& \int_0^\infty \rmd\omega\, \omega \, |\phi(\omega)|^2 =: \Ho (\phi) 
\ea
where, in the first step, we could extend the initial integral over the interval $R \in(0,R_\circ(T))$ to $R \in (0, \infty)$ because $\Phi(R,T)$ vanishes for $R > R_\circ(T)$. Note that $\Ho (\phi)$ is a constant because it equals the total, conserved energy of the solution $\Phi$ w.r.t. the Minkowski metric $\go_{ab}$; it is independent of $T$. Therefore, on $M^\prime$ the metric assumes the form 
\be \label{g} g_{ab}\,\rmd x^a\, \rmd x^b\mid_{M^\prime}\,\, =\, e^{G\Ho} \, (\rmd T^2 + \rmd R^2)\, +\, R^2 \rmd \theta^2\, \ee
showing that it is flat! Indeed, this is also obvious from the fact that Ricci tensor $R_{ab}$ vanishes in absence of sources and in 3-dimensions $R_{abcd}$ is determined entirely by $R_{ab}$. However, unless $\Ho=0$, --i.e., unless $\Phi$ vanishes identically-- the physical metric $g_{ab}$ is not globally flat: Although it is everywhere smooth on all of $M$, it has a `conical structure with a deficit angle' in $M^\prime$.\, Note also that $g_{ab}$ admits, a time translation Killing vector $\partial/\partial_T$ on $M^\prime$, in addition to the rotational Killing vector. The solution is asymptotically flat both at spatial and null infinity \cite{aamv,abs1,abs2}. 

Hamiltonian analysis of this midi-superspace \cite{aamv,aamp} reveals a striking result: The Hamiltonian $H(g,\Phi)$ --i.e., the generator of the asymptotic  time translation for the full system-- is \emph{not} the energy $\Ho(\phi)$ of the Klein Gordon field in Minkowski space, but is \emph{non-polynomially} related to it:
\be \label{H} H(g,\Phi) = \f{1}{4G}\, \big(1 - e^{- 4G\Ho(\phi)}\big) \ee
This is the physical energy of the system in the Einstein-Klein-Gordon theory; as with the ADM energy in 4-d, it includes all contributions, including those from gravity.%
\footnote{While the Hamiltonian analysis refers to spatial infinity, $H$ is also the past limit of the Bondi energy at null infinity \cite{abs1}.}
$H$ is manifestly positive and, in striking contrast to $\Ho$, it is \emph{bounded above}. This property is \emph{genuinely non-perturbative}: if we were to expand $H$ in powers of $G$, to the leading order $G^{\,0}$, we obtain $\Ho$ and if we truncate the series to any finite order, the expression is again unbounded above. It is only when one sums all the terms that one finds that $H$ is bounded. For a flat conic space-time, this result was obtained by Henneaux already in 1983 \cite{henneaux} and, for geometries sourced by $N$ point particles, by Deser, Jackiw and 't Hooft \cite{djt} in 1984. But in fact positivity and boundedness of $H$ is a general result in 3-d general relativity, so long as the matter satisfies the positive energy condition $T_{ab}\, t^a_1\, t^b_2 \ge 0$ for all time-like vectors $t^a_1, t^a_2$ \cite{aamv}. Finally, note that the non-perturbative form (\ref{H}) is possible because the 3-d Newton's constant has physical dimensions of inverse mass.

In 4-d gravity the Arnowitt-Deser-Misner (ADM) or the Bondi energy is expressible as a 2-surface `charge-integral' at infinity that involves only geometrical/gravitational fields. In the 3-d case the expression (\ref{H}) of energy, on the other hand, involves $\Ho$  which is a bulk integral involving just the matter fields. However, since it  measures the deficit angle of the metric $g_{ab}$ in $M^\prime$, it has a geometric interpretation (see \cite{djt} for geometries sourced by point particles). Moreover, as we now  explain, it can also be expressed as a \emph{line integral involving only geometry}. Since the isometry group in the tangent space of any point of $(M, g_{ab})$ is the 3-d Lorentz group, there is a natural notion of $SU(1,1)$ spinors, $\lambda_A$. By parallel transporting any $\lambda_A$ at a point $p\in M$ along a closed curve $C$ starting and ending at $p$ yields another spinor $\lambda^\prime_A$ at $p$, related to the original $\lambda_A$ via $\lambda^\prime_A = U^{(C)}{}_A{}^B\,(p)\, \lambda_B$, where $U^{(C)}{}_A{}^B\,(p)$ is an $SU(1,1)$ rotation. Now, ${\rm Tr}\, U^{(C)}$ =  $U^{(C)}{}_A{}^A (p)$  is independent of the choice of the point $p$ on the curve. Furthermore, since the curvature of $g_{ab}$ vanishes in $M^\prime$, ${\rm Tr}\, U^{(C)}$ is also independent of the choice of the curve $C$ that lies entirely in $M^\prime$. Let us restrict ourselves to such curves. Then, the trace is a function only of the total energy $H(g,\Phi)$ of the system:
\be \label{hol} {\rm Tr}\, U^{(C)}(g) = 2 \cos \Big[\f{\pi H(g,\phi)}{4G}\Big] \, .  \ee
Consequently, there is a simple thought experiment that we can make in the asymptotic region to determine $H$, and therefore also the metric coefficient $g_{RR}$ (or $G_{TT}$) in the source-free region $M^\prime$: transport particles with spin\, $\f{1}{2}$\, around any closed loop $C$ in $M^\prime$, measure the $SU(1,1)$ rotation $U_C(p)$ they undergo, and take its trace. Since $\Phi$ vanishes on $M^\prime$, this experiment can be carried out \emph{without  reference to matter fields}. We will return to this experiment in section \ref{s4}.

Finally, there is another striking feature of 3-d gravity: the absence of gravitational collapse leading to the formation of black holes. This is related to a fact we noted above: in 3-d there is no analog of the Schwarzschild radius associated with a given mass $M_\circ$. In our midi-superspace, this feature is  seen in detail: For regular initial data for $\Phi$ (e.g. of compact support), no matter how densely we pack the scalar field, we obtain a non-singular solution $g_{ab}$ on all of $M = {R}^3$ without any horizons. 

We conclude with a number of remarks addressed primarily to the general relativity community.\medskip

\emph{Remarks:} 

(1) An important difference in the asymptotic structure of 4-d and 3-d gravity is the following. In 4-d  we can and do ask that all physical metrics $g_{ab}$ of interest approach a \emph{fixed} Minkowski metric $\go_{ab}$ asymptotically and the information about the mass is encoded in the leading deviation. In 3-d, as we just saw, $g_{ab}$ does not approach the fixed Minkowski metric even to the zeroth order because of the factor involving $\Ho$ that varies from one space-time to another, depending on its energy content. This fact complicates the discussion of the asymptotic symmetry group in both spatial \cite{aamv} and null \cite{abs1} regimes. In particular, the notion of `time translation' is now more subtle. This issue of time is discussed in, e.g., \cite{aamp,mena}.

(2) The model we presented is a symmetry reduction of the celebrated Einstein-Rosen waves of 4-d, vacuum general relativity \cite{EinRos,exact}. These solutions have cylindrical symmetry --i.e., two hypersurface orthogonal, commuting Killing vector fields, a spatial translation (say along direction $z$ direction) and a rotation (around the $z$ axis). The Kaluza Klein reduction is carried about for the translational Killing field with norm, say $e^{\psi}$ and the 3-d scalar field $\Phi$ is given by: $\Phi = \psi/4\pi G$ \cite{abs1}. The rotational Killing field of ${}^4\!g_{ab}$ descends from the 4-manifold ${}^4\!M$ to the 3-manifold $M$ and becomes a Killing field of the 3-metric $g_{ab}$. Einstein's vacuum equations ${}^4\!R_{ab} =0$ in 4-d are equivalent to the 3-dimensional field equations (\ref{fe}). For details, especially of the asymptotic analysis from a 4-d perspective, see \cite{abs2}. In the context of Einstein-Rosen waves, $\Ho$ is called the `c-energy' and was sometimes interpreted as the energy `per unit length along the $z$-axis' in these waves. Careful Hamiltonian analysis \cite{aamv,aamp} shows that it is more appropriate to assign this interpretation to $H$.

(3) There is an interesting variation \cite{schmidt} of the midi-superspace of 4-d Einstein-Rosen waves in which the space-time topology is $R^2 \times T^2$ rather than $R^4$ and, consequently, the corresponding $\mathcal{I}^+$ has topology $T^2\times R$.  
In the Kaluza-Klein reduction, the 3-d space-time has topology $R^2\times S^1$, rather than $R^3$. Locally, the structure of equations is the same as the one we discussed but there are global differences. In particular, there is a right and a left spatial infinity. This model has been analyzed from a Hamiltonian perspective in detail and then used as a point of departure for non-perturbative quantization in \cite{beetle}. The canonical analysis along the same lines as \cite{aamp} but with interesting conceptual differences that arise from the difference in boundary conditions. Another interesting exactly soluble model is provided by Gowdy space-times, where the spatial topology is that of a 3-torus $T^3$. Here the midi-superspace is significantly richer in that the 4-d  Killing vector used in the Kaluza-Klein reduction is not hypersurface orthogonal. Consequently, there is an additional local degree of freedom in the 3-dimensional description. The phase space and non-perturbative canonical quantization of this model is discussed in \cite{mena} in the setting of 4-dimensional connection-dynamics.

\section{Non-perturbative quantum theory}
\label{s3}

This section is divided into three parts. In the first we introduce the quantum framework, in the second we discuss the unforeseen results on the nature of quantum geometry that it leads to, and in the third we discuss a thought experiment  that brings out the deep interplay between the notion of physical reality and the theoretical paradigm used to frame it in.

\subsection{The framework}
\label{s3.1}

Classically the model is exactly soluble because we can decouple the matter field $\Phi$ and the dynamical space-time metric $g_{ab}$. This procedure led us to encode the unconstrained degree of freedom of the total system in $\Phi$ satisfying $\Boxo\,\Phi =0$, or, equivalently, in the freely specifiable function $\phi(\omega)$. Therefore, a natural strategy is to first focus on $\Phi$ and quantize it and then investigate the nature of the quantum geometry $\hat{g}_{ab}$ it determines. This procedure will make the underlying mathematical structure simple since the first step involves just quantum fields in 3-d Minkowski space-time $(M,\go_{ab})$. 

Let us first collect the equations we will need from the textbook quantization of the Klein-Gordon field $\Phi(R,T)$ in the 3-d Minkowski space-time $(M, \go_{ab})$. The positive and negative frequency expansion (\ref{sol2}) of the solution immediately leads us to the `field operator'
\be \label{ovd} \hat\Phi(R,T) = \int_0^\infty \rmd \omega\, \big[ \phi^+_\omega (R,T)\, \ha +\,  \overline{\phi^+}_\omega (R,T)\, \had \big]\,  \ee
where $\ha$ and $\had$ are the annihilation and creation operators satisfying $[\ha\, \hat{A}^\dag(\omega^\prime)] = \hbar\, \delta(\omega, \omega^\prime)$. They operate on the symmetric Fock space $\F$, where the underlying 1-particle Hilbert space $\H$ is spanned by $\phi(\omega)$ --or, equivalently, classical solutions $\Phi(R,T)$-- \emph{which have a finite norm}:
\be \label{norm}  ||\phi||^2 := \f{1}{\hbar} \, \int_0^\infty\! \rmd\omega\, |\phi(\omega)|^2\, .  \ee
Note that, while solutions $\Phi(R,T)$ and their `frequency components'  $\phi(\omega)$ are classical concepts that make no reference to $\hbar$, presence of $\hbar$ is essential in the expression (\ref{norm}) since the norm has to be dimensionless; (\ref{norm}) is a quintessentially quantum expression. The Hamiltonian operator $\hat\Ho$ has the standard expression
\be \label{Ho} \hat\Ho = \int_0^\infty \rmd\omega\, \omega \had\, \ha \, . \ee
(There is no explicit factor of $\hbar$ because $[\ha,\, \hat{A}^\dag(\o^\prime)] = \hbar\,\delta(\omega,\omega^\prime)$.)

As is well-known, in the free field theory coherent states play a key role in the discussion of the relation the quantum theory to classical. Given a classical solution $\mathring\Phi(R,T)$, or equivalently, $\phio(\o)$, we have a normalized coherent state $\kcoh$:
\ba \label{coherent} \kcoh &=& N\,\, e^{\f{1}{\hbar}\int \rmd \o\, \phio(\o) \had}\, |0\rangle\nonumber\\  {\rm where}\quad N &=& e^{\f{1}{2\hbar}\, \int \rmd \omega \, |\phio(\o)|^2}\quad {\hbox {\rm is the normalization constant.}}\ea
Recall that the expectation values of $\h\Phi(R,T)$ and $\h{H}$ in this state just yield the classical solution $\mathring{\Phi}(R,T)$ and its energy:
\be \bcoh \h\Phi(R,T) \kcoh\, =\, \Phi_\circ(R,T) \qquad {\rm and} \qquad \bcoh \h{\Ho}(R,T) \kcoh\, =\, \Ho (\phio)\,. \ee
Furthermore, as is well-known, the state $\kcoh$ is sharply peaked at the classical solution $\Phi_\circ(R,T)$ in the sense that the uncertainty in the field and its momentum is saturated and equally distributed in the appropriate sense at all times $T$. The uncertainty in the Hamiltonian $\Ho$ is given by:
\be \label{hofluct} \big(\Delta\hat{\Ho} \big)^2 = \hbar \int_0^\infty \rmd \omega\, \omega^2\, |\phio(\o)|^2 \quad {\rm and} \quad
\f{\big(\Delta\hat{\Ho} \big)}{\big(\langle \hat{H} \rangle\big)}\,  \approx\,  \f{1}{\big(\f{1}{\hbar}\int_0^\infty \rmd \omega\, |\phio|^2(\o)\big)^{\f{1}{2}}} \, =\, \f{1}{\langle \hat{\mathcal{N}} \rangle^{\f{1}{2}}}\, . \ee
In the second equation we have considered a profile $\phio(\o)$ that is sharply peaked at some fixed frequency to bring out the physical meaning of the right side, and denoted the number operator on the Fock space $\F$ by $\hat{\mathcal{N}}$. Thus, the relative uncertainty is proportional to the inverse of the square root of the expected number of `photons' in the coherent state under consideration. More intense the classical beam $\phio(\o)$ --i.e., greater the value of $\int \rmd \omega\, |\phio|^2(\o)$ relative to $\hbar$-- less the relative uncertainty in the quantum state, and more trustworthy are the classical results vis a vis the correct quantum answer. Note that the frequency at which $\phio(\o)$ is peaked is irrelevant in this consideration --we could shift the peak to arbitrarily high frequencies and the classical approximation will continue to be excellent so long as $\int_0^\infty \rmd \omega\, |\phio|^2(\o)\, \gg \hbar$. This is a standard result in quantum optics. 

But all these considerations referred only to the matter sector --either a scalar field $\Phi$, or a Maxwell field $F_{ab} = \epsilon_{abc}\nabla^c\Phi$-- which, thanks to the integrability of the model, could be analyzed without knowing the physical metric $g_{ab}$. Now, in the classical theory, Einstein's equations determine $g_{ab}$ completely for any given matter field $\mathring\Phi(R,T)$.  %via (\ref{g}) and (\ref{sol1}). 
In the quantum theory, we have chosen to put the true degree of freedom in $\Phi(R,T)$ and represented it as an operator (\ref{ovd}) on $\F$. Therefore, the metric coefficient $g_{RR} = -g_{TT}$ in the asymptotic region --namely, $e^{G\Ho}$ of (\ref{g})--  also becomes an operator on $\F$:

\be \label{gq} \hat{g}_{RR}\mid_{\rm asym}\, :=\, e^{G\hat{\Ho}}\, =\, e^{G\int_0^\infty \rmd \o\, \o\, \had\, \ha}\, . \ee
Similarly, the non-perturbative  Hamiltonian $H$ of the total system --matter \emph{and} gravity-- becomes the operator
\be \label{Hq} \hat{H}\, := \, \f{1}{4G}\, \big( 1 - e^{-4G\hat{\Ho}}\big)\,  \ee
on $\F$. These operators encode the non-perturbative information contained in gravity-matter interactions. 

Generally non-trivial models are exactly soluble because they can be mapped to a trivial model -- in our case, the Einstein-Klein-Gordon (or Einstein-Maxwell) system $(g_{ab}, \Phi)$ could be mapped to a free field $\Phi(R,T)$ in Minkowski space. Non-triviality is then transferred to the map that \emph{relates the two models}. In our case, the relation is given by (\ref{g}) and (\ref{H}) in the classical theory and (\ref{gq}) and (\ref{Hq}) in the quantum theory. In section \ref{s3.2} we will find that these equations have certain unforeseen consequences that bring out the non-triviality of the original Einstein-matter system.\medskip 

\emph{Remarks:}

1. While defining various operators one encounters the issue of factor ordering. For the free Hamiltonian $\hat{\Ho}$ of the scalar field $\hat\Phi$ in Minkowski space, there is an unambiguous answer: normal ordering used in (\ref{Ho}). In view of the relation (\ref{g}) between the classical $\Ho$ and metric coefficients $g_{RR} = -g_{TT}$, and the relation (\ref{H}) between $\Ho$ and the Hamiltonian $H$ of the full system then led us to define quantum operators 
$\h{g}_{RR}|_{\rm asym}$ and $\hat{H}$. However, it is natural to ask whether one could also use other choices. 
For example, could we not have normal ordered the operator \emph{after} exponentiation and used $\hat{H}^\prime\, =\,\, :\!\hat{H}\!:$ instead? The operator $\hat{H}^\prime$ shares several attractive properties with $\hat{H}$. In particular, it also annihilates the vacuum state $|0\rangle \in \F$, and its expectation value in coherent states $\kcoh$\, also yields the value of the classical Hamiltonian $H(\phio)$. However, in sharp contrast to $\hat{H}$, the spectrum of $\hat{H}^\prime$ is the \emph{entire real line}. Given that the classical observable $H$ takes values only in the finite interval $[0,\, \f{1}{4G}]$,\,\, $\hat{H}^\prime$ is an inadmissible choice for the corresponding quantum observable. The choice we made in (\ref{gq}) and (\ref{Hq}) is free of such drawbacks. In particular, the spectrum of $\hat{H}$ is precisely $[0,\, \f{1}{4G}]$.

2. The `field operator' $\hat\Phi(R,T)$ of (\ref{ovd}) is in fact an operator-valued distribution. Consequently, there are technical subtleties in giving precise mathematical meaning to $\hat{g}_{RR} = - \hat{g}_{TT}$ in full space-time (discussed,  e.g., in \cite{aamp}). However, for conceptual issues that are at the forefront of our analysis, we could bypass this issue by focusing only on the asymptotic region, where the classical matter fields $\Phi_{\circ}(R,T)$ --at which the coherent states $\kcoh$ are peaked-- vanish. More precisely, the careful treatment yields the same results as those we obtained by restricting the discussion to the asymptotic region from the beginning.
 
3. In our framework the true degree of freedom lies in the scalar Field $\Phi(R,T)$. Therefore, it was natural to construct the Hilbert space of states for this field and represent geometric observables by appropriate operators thereon. However, one might be concerned that this procedure is somewhat unnatural from the perspective of general relativity and \emph{only} the physical metric $g_{ab}$ should be used in the quantization procedure. This is indeed possible using a canonical quantization procedure which refers only to the physical metric as was first discussed in the 4-d context of cylindrically symmetric waves in a remarkably early work by Kucha\v{r} \cite{kuchar}, and later in the 3-d context in \cite{allen} (without however realizing that the second is a Kaluza-Klein reduction of the first). But these investigations overlooked subtle issues related to boundary conditions and the distinction between diffeomorphisms representing gauge and true dynamics they imply. A careful handling of these issues is needed to put the Hamiltonian theory and the canonical quantization procedure on a firm footing. When this is done, the framework also brings out subtleties associated with the issue of time. Finally, the Hilbert space ($\F$ in our discussion above) of quantum states can also be selected without having to introduce the Minkowski metric $\go_{ab}$. For details on both these points, see \cite{aamp}.

\subsection{Unforeseen quantum gravity effects}
\label{s3.2}

The Hilbert space of quantum states is the Fock space $\F$, constructed entirely from the matter sector.  As we saw, if one uses observables --such as $\hat\Phi$ and $\hat{\Ho}$-- that refer only to the matter sector, then the classical theory is an excellent approximation if the system is in a coherent state $\kcoh$. In these states, the expectation values of quantum observables equal values of their classical counterparts and their  quantum fluctuations are negligible provided the expected number of `photons', $\bcoh \hat{\N}\kcoh$, is large, irrespective of the choice of $\phio$. Now, observables --such as $\hat{g}_{RR}$ that encodes the quantum metric and $\hat{H}$ that represents the Hamiltonian $\hat{H}$ of the full system, including gravity-- are also represented by operators on $\F$. In the classical theory, values of these observables can be computed using the matter sector alone. Therefore one's first expectation would be that the classical theory would again provide an excellent approximation for these observables if the system is in a coherent state $\kcoh$. However, as we will now show, this expectation is not borne out. Even though geometry is completely determined by matter and coherent states $\kcoh$ are sharply peaked on classical configurations $\phio$, quantum properties of gravitational observables in these states can be very different from their classical analogs. For notational simplicity, from now on we will drop the suffix `asymp' in $g_{RR}\!\mid_{\rm asym}$ since our discussion will refer only to the asymptotic metric.

Let us begin with the action of the operator $\hat{g}_{RR} = e^{G\hat{\Ho}}$ on coherent states:
\ba \label{long} e^{G\hat{\Ho}} \kcoh\, &=& N\, e^{G\hat{\Ho}} \, e^{\f{1}{\hbar}\int_0^\infty\, \rmd \o\, \phio(\o)\, \had }\,\,|0\rangle\nonumber\\
&=& N\, e^{G\hat{\Ho}} \, \sum_{n=0}^\infty \, \f{1}{\hbar^n\, n!}\, \int_0^\infty\,\rmd \o_1 \ldots \rmd \o_n\,\, \phio(\o_1) \ldots \phi(\o_n)\,\,|\o_1, \ldots \o_n\rangle \nonumber \\
&=& N\, \sum_{n=0}^\infty \,\f{1}{\hbar^n\, n!}\, \int_0^\infty\,\rmd \o_1 \ldots \rmd \o_n\,\, \phio(\o_1) \ldots \phi(\o_n)\,\, e^{G\hbar(\o_1 + \ldots +\o_n)}\,\, |\o_1, \ldots \o_n\rangle \nonumber\\ 
&=& N\, \sum_{n=0}^\infty \,\f{1}{\hbar^n\, n!}\, \int_0^\infty\,\rmd \o_1 \ldots \rmd \o_n\,\, \bfphi(\o_1) \ldots \bfphi(\o_n)\,\, \,\, |\o_1, \ldots \o_n\rangle\, , 
\ea
where, as before,  $N = e^{\f{1}{2\hbar}\,\int \rmd \o\, |\phio(\o)|^2}$ is the normalization constant for the coherent state, and in the last step we have set  $\bfphi (\o) = e^{G\hbar\o}\,\phio (\o)$. Using the definition of the normalized coherent state $|\Psi_{\bfphi}\rangle$, we can express the result as:
\be \label{final}  e^{G\hat{\Ho}}\kcoh\, = \,e^{\f{1}{2\hbar}\int_0^\infty \rmd\o\, |\phio (\o)|^2 (e^{2G\hbar\omega} -1)}\, 
|\Psi_{\bfphi}\rangle \, . \ee
Thus, the action \emph{shifts} the peak of the coherent state from $\phio$ to $\bfphi$ and multiplies it by a constant. Note that the shift rescales $\phio(\o)$ by a factor that is \emph{exponential} in $G\hbar\omega$. Finally, since the inner-product between these two coherent states is given by
\be \bcoh \Psi_{\bfphi}\rangle\, =\, e^{-\f{1}{2\hbar}\, \int_0^\infty \rmd\o |\phio(\o)|^2 (1+ e^{2G\hbar\o} -2 e^{G\hbar\o})} \ee
the expectation value of $\hat{g}_{RR}|_{\rm asym}$ is given by
\be \label{exp1} \bcoh\, \hat{g}_{RR}\, \kcoh \, =\, e^{\f{1}{\hbar}\int_0^\infty \rmd\o\, |\phio(\o)|^2\,(e^{G\hbar\omega} -1)}\, , \ee
in contrast to the classical value
\be \label{class1} g_{RR} = e^{G \int_0^\infty \rmd\o\, \o\, |\phio(\o)|^2 } \, .\ee

There are several notable differences between the expectation value (\ref{exp1}) and the classical value (\ref{class1}). First, even though the expectation value is evaluated in a coherent state, it depends explicitly on $\hbar$. Thus, unlike  $\bcoh \hat\Phi(R,T) \kcoh$ or $\bcoh \hat{\Ho} \kcoh$, in the matter sector, in the geometric sector the expectation value (\ref{exp1}) itself carries a signature of quantum effects. Second, since we only had $\hbar$ at our disposal in the matter sector, we did not have a preferred frequency scale. With the availability of both $G$ and $\hbar$, we do: The Planck frequency is given by $\omegap = (G\hbar)^{-1}$. Therefore we can examine the expectation value in various limits by assuming that $\phio(\o)$ is sharply peaked at various  frequencies $\omegao$. Let us first consider the low frequency limit, $G\hbar\omegao \ll 1$. Then, the expectation value (\ref{exp1}) can be approximated as:
\be \label{low1} \bcoh\, \hat{g}_{RR}\, \kcoh \,\, \approx\,\, e^{G \int_0^\infty \rmd\o\, \o\, |\phio(\o)|^2}\,\, \big(1\, +\N\, (G\hbar\omegao)^2 \big)\,. \ee
where as before $\N$ is the expected number of `photons' in the state $\kcoh$. Thus, we recover the classical value of $g_{RR}$ \emph{provided the frequency is so low that} $\N\, G\hbar\omegao \ll 1$. Interestingly, while in the matter sector the classical approximation becomes better as $\N$ increases, for geometry, it becomes worse as $\N$ increases even when $\omegao$ is in the low frequency regime!%
\footnote{I thank Don Marolf for pointing out that in hindsight this second condition can be understood by first noting the relation $g_{RR} = e^{G\Ho}$ and then using the form (\ref{hofluct}) of fluctuations in $\hat{\Ho}$. Although the result refers only to to the expectation value of $\hat{g}_{RR}$ and not to fluctuations, the argument makes the disagreement between classical expectations and exact quantum results plausible.} 
In the high frequency regime, $G\hbar\omegao \gg 1$, there is \emph{always a huge disagreement} between the expectation value and the classical expression:  
\be \label{high1} \bcoh\, \hat{g}_{RR}\, \kcoh \,\, \approx\,\, e^{\N (e^{G\hbar\omegao})} \quad \hbox{\rm so that}\quad 
\f{\bcoh\, \hat{g}_{RR}\, \kcoh}{g_{RR}} \, \approx\, e^{\N\, ( e^{G\hbar\omegao}\, - \,G\hbar\omegao)}\, .
\ee
In this regime, $\hbar$ does not disappear; in fact terms involving $\hbar$ swamp the classical term since in the numerator $G\hbar\omegao$\, appears \emph{exponentially} in exponent itself! And again, larger the expected number of `photons' in the source, larger is the disagreement between the quantum and the classical results. 

We can also calculate the quantum fluctuations. In the matter sector, these are very small provided $\N \gg 1$, irrespective of the frequency at which $\phio(\o)$ is concentrated. The situation is again quite different for quantum geometry. Since $\hat{G}_{RR}^2 = e^{2G \hat{\Ho}}$ the same calculation we carried out to evaluate $\bcoh\, \hat{g}_{RR}\, \kcoh$ yields:
\be \bcoh\, \hat{g}_{RR}^2\, \kcoh \, =\, e^{\f{1}{\hbar}\int_0^\infty \rmd\o\, |\phio(\o)|^2\,(e^{2G\hbar\omega} -1)}\, , \ee
whence the relative uncertainty is given by
\be \Big(\frac{\triangle\hat{g}_{RR}}{\langle\hat{g}_{RR} \rangle} \Big)^2\,\,
= \,\, [e^{{\frac{1}{\hbar}}\int \rmd\o\, |\phio(\omega)|^2 (1 - e^{G\hbar\o})^2}
\,\,\, -\, 1]\, . \ee
We can again simplify this exact expression using $\phio(\o)$ that is sharply peaked at some frequency $\omegao$. Then, in the low frequency regime with $G\hbar\omegao \ll 1$ we obtain
\be \label{low2} \Big(\frac{\triangle \hat{g}_{RR}}{\langle\hat{g}_{RR}\rangle}\Big)^2 \,\,\approx\,\,
e^{\N (G\hbar\omegao )^2}\,\, - 1\ ,
\ee
Again, we find that the uncertainties in geometry grow with the number $\N$ of `photons'. Thus, there is an interesting tension: To reduce quantum fluctuations in the matter sector, we need a large number $\N$ of `photons' in the coherent state $\kcoh$. However, for a fixed $\omegao$, larger the value of $\N$ --i.e., more intense the beam-- larger is its influence on gravity. In quantum theory, this gives rise to larger relative uncertainties. The qualitative nature of this effect is not surprising. Nonetheless, it is pleasing to see it appear in a sharp, precise and quantitative form. This is possible because the model is exactly soluble. 

In the high frequency regime, as one would expect from our results on expectation values, quantum fluctuations in geometry are huge
\be \label{high2} \Big(\frac{\triangle\hat{g}_{RR}}{\langle\hat{g}_{RR}\rangle}\Big)^2 \,\,\approx\,\, e^{\N\, (e^{2G\hbar\o_o})}\, 
\ee
since $G\hbar \o \gg 1$ and appears \emph{exponentially} in the first exponent. Again, these effects refer to the asymptotic form of the metric in a region in which the classical profile $\phio$ can be vanishingly small. In the classical theory the metric in this region is as tame as it can be since the curvature vanishes. Yet quantum geometry exhibits very large, quintessentially quantum mechanical effects even though it has no degrees of freedom of its own and is completely determined by the matter sector!

Finally, for the total Hamiltonian $\hat{H}$ of the matter+gravity system, the analysis can be carried out using the same techniques. For completeness, let us list the final results. The expectation value of $\hat{H}$ is given by:
\be \bcoh \hat{H}\kcoh\, =\, {\frac{1}{4G}}\,\,\big[1 - e^{{\frac{1}{\hbar}} 
\int_0^\infty \rmd\o\, |\phio(\o )|^2 (e^{-4G\hbar\o} -1)}\big]\,, \ee
while the value of the classical Hamiltonian $H(\phio)$ is
\be H(\phio) = \f{1}{4G}\, \big( 1- e^{-4G\int_0^\infty \rmd \o |\phio(\o)|^2} \big)\, . \ee
In the low frequency regime, $G\hbar\omegao \ll 1$ we have
\be \bcoh \hat{H} \kcoh \approx  H(\phio) - \frac{4}{G}\N (G\hbar \omegao)^2 
e^{-4 G H_o (\phio)}\, , \ee 
while in the high frequency limit $G\hbar \omegao \gg 1$ we have:
\be \bcoh \hat{H} \kcoh \approx  \f{1}{4G}\, \big(1 - e^{-\f{\Ho(\phio)}{\hbar \omegao}}\big)\ee
which is quite different from $\Ho(\phio)$! The uncertainties can also be calculated exactly. But the significance of quantum effects is is now overshadowed by the fact that the spectrum of $\hat{H}$ is bounded in the finite interval $[0,\, \f{1}{4G}]$.

Let us summarize. One of our primary motivations is to analyze the quantum effects on geometry induced by quantum matter. Therefore, we are led to ask: Can matter induce interesting quantum effects on the geometrical/gravitational sectors, even when it is itself in `tame' quantum states? This question led us to focus on standard coherent states in the matter sector because they provide canonical `quantum representations of classical matter fields'. And we found that the quantum effects induced on geometry/gravity can be very large even when quantum fluctuations in the matter sector are small. This unforeseen behavior arises because of the non-linear structure that is specific to Einstein dynamics: These non-linearities can magnify small fluctuations in the matter sector to huge quantum effects on geometry/gravity. Not surprisingly, the effect is very pronounced when the classical configuration $\phio(\o)$ is peaked at a trans-Planckian frequency $\omegao \gg \omegap =\f{1}{\hbar G}$. Then the expectation value  $\bcoh\, \hat{g}_{RR}\, \kcoh$ is wildly different from the classical $g_{RR}(\phio)$, even in the asymptotic region where $\phio$ vanishes. But there are interesting effects also in the `tame' sub-Planckian frequency regime $G\hbar\omegao \ll 1$, because what enters in the explicit expressions of quantum effects in the gravitational sector is $\sqrt{\N} \times G\hbar \omegao$ --rather than just $G\hbar \omegao$-- where $\N$ is the expected number of `photons' in the coherent state $\kcoh$. In the matter sector, classical theory provides better and better approximation to the full quantum theory as $\sqrt{\N}$ increases. Therefore, it is also natural to consider the limit in which $G\hbar \omegao \ll 1$ but $\sqrt{\N}\,  (G\hbar \omegao) >1$. In this regime, the matter sector is extremely well approximated by the classical theory. But (\ref{low1}) implies that $\bcoh\,\hat{g}_{RR}\,\kcoh$  will not be well approximated by the $g_{RR}$ of the classical solution.
%$\bcoh\,\hat{g}_{RR}\,\kcoh$ will be more than 1.6 times as large as the classical value $g_{RR}(\phio)$, the difference between the two being proportional $\hbar$. 
Furthermore, (\ref{low2}) implies that in this regime quantum fluctuations will be small also in the geometrical sector. Thus, the state $\kcoh$ will be peaked on a classical geometry but different from the one determined by classical source $\phio(\o) \sim \mathring\Phi (R,T)$!

Finally, since $\hat{g}_{RR}$ is a (positive definite) self-adjoint operator on $\F$, we can use its spectral decomposition to  construct states in $\F$ that are sharply peaked at a classical value $g_{RR}(\phio)$. What happens to the surprising quantum effects in these states? This interesting issue is discussed in \cite{rgjp}. It turns out that the large quantum effects are then transferred to the matter sector. We chose to focus on $\kcoh$ because they are better suited to analyze quantum effects on geometry \emph{induced by matter}: States in which matter observables have tame behavior are best suited to bring out this induced effect.

\emph{Remark:} In our analysis, we began with the standard description of general relativity in terms of the metric. Now, in the asymptotically flat context, there exists in the literature an alternate formulation of Einstein's equations where the basic variable is not a metric $g_{ab}$ but a `cut-function' $Z$ that, at the end of the analysis, determines $g_{ab}$. In the final picture, $Z$ specifies intersections of light cones, emanating from space-time points, with $\mathcal{I}^+$ \cite{NSF1}. The underlying idea was to pass to quantum general relativity via an appropriate quantization of $Z$ --rather than $g_{ab}$-- with the hope that in the final theory space-time points themselves would become fuzzy \cite{NSF2,NSF3}. In the 2+1 theory under consideration, the metric $g_{ab}(R,T)$ is replaced by a single function $Z(\xi; R,T)$ where $(u, \xi)$ label points of $\mathcal{I}^+$ and in the final picture the 3-parameter family of `cuts' defined by space-time points is given by $u = Z$.
In the 3+1 theory, it has not been possible to implement this idea beyond linearized approximation. However, it could be carried out non-perturbatively in the 2+1 theory, using the framework outlined in this section as point of departure \cite{admt}. Interestingly, it was found that, even though there are large quantum fluctuations in the metric, fluctuations in $\hat{Z}$ are strongly damped. In this framework there are dynamical variables that correspond to space-time points and their fluctuations are similar to those of $\hat{g}_{RR}$ we found. In this sense, the idea of `fuzzy points' is realized in an interesting manner: space-time points can lave large fluctuations but fluctuations of the intersections of their light cones with $\mathcal{I}^+$ are highly suppressed.

\subsection{Lessons from a thought experiment}
\label{s3.3}

As explained in section \ref{s1}, exactly soluble models are often well suited to sharpen conceptual issues. One such issue is: how does one's description of physical reality change when one passes from a less accurate theory to a more accurate one? General relativity, for example, opened entirely new classes of phenomena and possibilities that could not be envisaged in Newtonian gravity. These arise because, with both $G$ and the velocity of light $c$ at one's disposal, a new scale arises. One can now associate a length with a mass --the Schwarzschild radius $R_{\rm Sch} = 2GM/c^2$ that is not available in the Newtonian theory since it knows only about $G$-- and this scale then unleashes an entirely new class of phenomena. Similarly, in the quantum theory Planck's constant $\hbar$ provides new scales, dramatically changing our understanding of the atomic world and leading to a plethora of unforeseen phenomena that have shaped physics of the micro-world. In quantum gravity one has access to all three of these fundamental constants and there have been speculations, dating back to Planck himself, on the nature of new physics that would arise at Planck length, Planck frequency, and Planck density. Exactly soluble models provide a clean-cut platform to discuss the nature of this new physics. In this subsection we will leave $c$ unchanged but switch on and off $\hbar$ and $G$ and examine how the nature of physical reality changes. 

Consider then a trivial thought experiment in a 3-dimensional space-time: Switch on a beam of laser light and then switch it off after some time.% 
\footnote{The laser beam would be naturally axisymmetric. We wish to focus just on the light beam and not details of the source that produced it. Therefore, we will consider as our system the retarded minus advanced solution, that satisfies the source-free Maxwell's equation.}
Our task is to describe what `really' happens in the space-time around us when this is done. Interestingly, we obtain four \emph{quite different} answers to this question: (i) A description of a \emph{classical physicist} who ignores both general relativity (GR) and quantum field theory (QFT) and sees only a Maxwell field propagating in flat space-time, setting $\hbar= G =0$;\, (ii) that of a \emph{quantum field theorist} who knows of $\hbar$ and describes the laser beam as a coherent state of photons but sets $G=0$;\, (iii) that of a \emph{general relativist} who ignores $\hbar$ but knows that the beam curves space-time with surprising consequences in the non-perturbative regime; and, finally, \, (iv) that of a \emph{quantum gravity theorist} who knows about both $G$ and $\hbar$. As our discussion of the last two subsection shows, the fourth description of physical reality has both subtle and truly novel elements. 

Let us begin with the first description. The physical system of interest for the classical physicist is the Maxwell field 
$\mathring{F}_{ab} = \epsilon_{abc}\nabla^c \mathring\Phi$ that propagates in a 3-d Minkowski  $(M, \,\go_{ab})$. It has initial data of compact support on every Cauchy surface and therefore vanishes in a finite neighborhood of $i^\circ$. The total energy in the system is entirely from the Maxwell field and equals $\Ho(\mathring{F})$, which can also be read-off at $\mathcal{I}$ (for detailed expressions, see, \cite{abs2}). It is unbounded above and increases linearly with the frequency and intensity (i.e., $|\phio(\omega)|^2$) of the beam. With only the velocity of light $c$ at one's disposal, there is no frequency (nor energy) scale in the theory. Therefore one cannot speak of low or high frequency (nor of low or high energy) fields and the general description holds for any frequency or intensity of the laser beam. The second description is that of a quantum field theorist. With availability of $\hbar$, they can regard the beam as consisting of photons.  They will point out that what the classical physicist referred to as a classical Maxwell field $\mathring{F}_{ab} \sim \mathring{\Phi}$ is in fact a coherent state $\kcoh$ and calculate the expected number $\N = \f{1}{\hbar}\, \int_0^\infty \rmd \o\, |\phio|^2$ of photons in it. They now point out that although the expectation value $\bcoh \,\hat{F}_{ab}(R,T)\,\kcoh$ of the Maxwell field operator agrees with what the classical physicist's $\mathring{F}_{ab}(R,T)$, there are quantum fluctuations that the classical physicist missed and these would become measurable if $\N$ were small, decaying as $1/\sqrt{\N}$. But in this description, since $G=0$, the physical space-time is again Minkowski space $(M, \go_{ab})$.

The third description is that of general relativist, who has access to $G$ and Einstein's equations, but not to $\hbar$. Therefore, they continue to describe the light beam using the classical Maxwell field. But their description of physical reality is quite different from that of the classical physicist. Now the physical metric is $g_{ab}$ rather than $\go_{ab}$. It is dynamical within the support of the beam. More interestingly, although it is flat outside the support, if one parallel transport a test spinning particle around a closed curve $C$ lying \emph{entirely} in this region, one finds that its state changes when it returns to the starting point. The change is encoded in the non-trivial holonomy $U^{(C)}$ of (\ref{hol}). Thus, by performing experiments with probes that never interact with the light beam, they can determine $\Ho$\, --whence also the non-trivial metric components $g_{RR}$ ($= - g_{TT})$-- \,and know that space-time is \emph{not} described by Minkowski metric $\go_{ab}$ even in the asymptotic region. Perhaps the most startling discovery they would make is that total energy $H$ of the system --including the gravitational contributions-- is \emph{bounded from above, irrespective of high high the frequency of the intensity of the laser beam is!} This qualitatively new phenomenon occurs because $G^{-1}$ has dimensions of mass that sets the scale for the upper limit. Furthermore, though a theoretical investigation, the general relativist would discover that this interesting fact about Nature cannot be captured using a perturbative expansion in $G$, no matter how high an order $n$ in $(GM)^n$ you truncate the theory. 

Finally, let us bring-in the quantum gravity expert. With both $G$ and $\hbar$ available, they can introduce a new notion of Planck frequency $\omegap = \f{1}{G\hbar}$. With this new scale available, not only can they design new experiments to probe aspects of physical reality that had remained unexplored, but also solve some mysteries that the other three were left with. They can design experiments to probe the quantum nature of geometry in the asymptotic region where there are no photons at all. Since, furthermore, there are no gravitons --or radiative degrees of freedom-- in the gravitational field, \emph{these are `clean' tests of the nature of quantum information in geometry induced entirely by the quantum properties of matter}. If they could create laser beams peaked at a trans-Planckian frequency $\omegao > 1/G\hbar$, they would discover that even in the asymptotic regions, $\bcoh \, \h{g}_{RR} \, \kcoh$ is \emph{very} different from $g_{RR}$ predicted by the classical theory. Much to the surprise of the general relativist, this can occur even when the amplitude of $\phio(\o)$ is chosen to be sufficiently small for the space-time curvature computed using classical Einstein's equations to be quite small everywhere. A second regime would surprise the quantum field theorist. Consider the case when the intensity of the beam is high so that $\N  \gg 1$. Then the quantum fluctuations in the matter sector are small and the quantum field theorist would expect the induced quantum fluctuations in the geometry would be at least as small, if not much smaller. We found that this is not the case; as $\N$ increases, the relative quantum uncertainties in $\hat{g}_{RR}$ grow both in the low and high frequency regimes (see (\ref{low2}) and (\ref{high2}) respectively). Finally, consider the regime $G\hbar\omegao \ll 1$ but $\N \times G\hbar\omegao >1$, discussed at the end of section \ref{s3.2}. Neither the quantum field theorist nor the general relativist would have a reason to suspect their descriptions because they would see nothing unusual in the matter sector. The quantum field theorist would only see a coherent state with a large number of photons, that is therefore well-approximated by a smooth, classical Maxwell field $\mathring{F}_{ab}$, and the general relativist would see the gravitational field $\mathring{F}_{ab}$ sources. Yet, both descriptions are inadequate. Had general relativist carried out a careful measure of $g_{RR}$ they would have found that their theoretical prediction is significantly different from the observed value. This would have been regarded as an anomaly by the community, analogous to that in the perihelion shift of mercury before the discovery of general relativity. Had the quantum field theorist carried out a careful measurement of the $g_{RR}$ and constructed a statistical distribution, they would have been puzzled that an irreducible minimum persists in the relative uncertainty, no matter how accurately they measure this (to them) `classical quantity'; somewhat like the puzzle experienced by Penzias and Wilson before they knew about the cosmic microwave background. Thus, the quantum gravity expert would be able to resolve the uncomfortable puzzles that the general relativity and quantum field theory communities had encountered by confronting their detailed theoretical calculations with careful measurements in specific regimes. 

\section{Discussion}
\label{s4}

In the last two sections we analyzed an exactly soluble sector of non-perturbative quantum gravity in 2+1 dimensions. Let us now explore the lessons it offers. In 3+1 quantum gravity, we have a number of paradigms and conjectures. Are they realized in our soluble sector? Do our results offer new hints or bring out unforeseen subtleties?

Let us begin with the list of illustrative examples of proposals from section \ref{s1}. The first proposal pertains the possibility of violation of the local Lorentz invariance due to trans-Planckian quantum effects. Thanks to the underlying simplicity of the sector, we could carry out explicit calculations. They brought to forefront the surprising role of trans-Planckian frequencies. If the matter profile $\phio(\o)$ at which the coherent state $\kcoh$ is peaked has even a small `blip' at a frequency $\omegao > \omegap$, the \emph{induced} quantum effects on the space-time geometry become large: even in the asymptotic region away from the support of the matter source, the classical value $g_{RR}$ is a poor approximation to the expectation value $\bcoh\, \hat{G}_{RR}\, \kcoh$ and the relative quantum fluctuations in $\hat{g}_{RR}$ are large.%
\footnote{Recall that, starting from section \ref{s3} we have dropped the suffix `asym' in $\hat{g}_{RR}|_{\rm asym}$.}
Thus, the intuition that trans-Planckian frequencies may change the paradigm suggested by classical GR and standard QFT is indeed borne out in a precise sense. Do these effects, then, lead to the violation of local Lorentz invariance? The frequency $\omegao$ does refer to the asymptotic time-translation Killing field, $\partial/\partial T$. However, the \emph{dynamics} does not break \emph{local} Lorentz invariance. The field equation satisfied by $\hat{\Phi}(R,T)$, for example, is fully covariant. In particular, there is no quantum gravity induced vector field in dynamical equations that can be the source of such a breakdown.  Next, let us consider the status of `holography' in this model. As in 3+1 dimensions, neither general relativity nor quantum field theory has a built-in length scale to define an ultraviolet regime. Quantum gravity does, and we found that this theory does depart radically from classical GR and QFT in the ultraviolet regime, $\omega > \omegap = \f{1}{G\hbar}$. Do these effects then force us to adopt a holographic view in the non-perturbative quantum description? In the classical theory, the true degrees of freedom are encoded in the matter field that resides in the 3-manifold $M$. We found that the situation is the same in non-perturbative quantum theory; the quantum scalar field continues to reside in the bulk as in any standard quantum field theory. Bulk degrees of freedom are not replaced by `surface degrees of freedom'. Thus, even though the model exhibits novel features in the Planck regime, they are realized within the standard bulk degrees of freedom of quantum field theory.  I should add that one \emph{can} construct the quantum Hilbert space $\F$ using just null infinity $\mathcal{I}^+$ \cite{abs1,aamp}. But this is not what is meant by `holography'. Indeed, one can also construct the standard Fock space of photons in 3+1 dimensions using just $\mathcal{I}^+$.

Next, recall that quantum geometry also has novel features already in the \emph{low} frequency regime. In particular, we found that many `seemingly tame' classical solutions are physically `spurious'. Consider, for simplicity, solutions  $(\mathring\Phi,\, g_{ab})$  for which $\phio$ is sharply peaked at a frequency $\omegao \ll \omegap$. It would seem natural to assume that such solutions are `tame' and should arise in the classical limit of the full quantum theory. That is, one would have expected the quantum theory to admit states $\kcoh$ that are sharply peaked at such solutions. However, we saw that for this to occur, \emph{two} inequalities have to be satisfied: $\N \gg 1$ and $\N (G\hbar\omegao)^2 \ll 1$. The first is required to ensure that $\kcoh$ is sharply peaked at $\mathring\Phi$ but the second inequality is surprising in that it requires that $\N$ cannot be `too large'! Why is it needed? Suppose $\N$ is large as per first inequality and $\N (G\hbar\omegao)^2 = K$. Then $\bcoh \hat{g}_{RR} \kcoh \approx  e^{\f{K}{2}}\, g_{RR}$. Thus, the classical answer would not be a good approximation even at the expectation value level unless $K \ll 1$. Consequently, a large class of apparently `tame' classical solutions are simply not realized as approximations to the quantum theory; from the more complete perspective of quantum gravity, they are spurious. This surprising limitation seems to re-enforce Wheeler's `space-time foam' idea that the quantum vacuum state would also be quite different from the perturbative vacuum that is peaked at Minkowski space. Is this expectation borne out? Recall that the perturbative vacuum $|0\rangle$ is the coherent state $\kcoh$ with $\phio=0$, and is annihilated by the matter Hamiltonian $\hat{\Ho}$. Now, the full Hamiltonian $\hat{H}$ and the asymptotic metric operator $\hat{g}_{RR}$ are both self-adjoint and well-defined functions of $\hat{\Ho}$. Therefore, $|0\rangle$ is an eigenstate of both, with eigenvalue $0$ for $\hat{H}$ and $1$ for $\hat{g}_{RR}$. Thus the perturbative vacuum $|0\rangle$ is in fact an eigenstate of the quantum metric operator with Minkowski metric $\go_{ab}$ as its eigenvalue. Contrary to one's first expectation, there is nothing unruly about the ground state! Finally, as we noted in section \ref{s2}, there is a positive energy theorem in classical 2+1 dimensional general relativity \cite{aamv}. It goes over to the quantum theory since $\hat{H}$ is a non-negative self-adjoint operator and the vacuum $|0\rangle$ is  the unique state that is annihilated by it. Thus, again we find that while several aspects of proposals for novel physics in quantum gravity are realized, closely related aspects that were assumed to follow are not. 

These considerations have implications also on semi-classical gravity, an approximation in which gravity is treated classically but matter quantum mechanically and the effect of quantum matter on the classical metric is computed using semi-classical Einstein's equations $G_{ab} = 8\pi G\, \langle\, \hat{T}_{ab}\,\rangle$, in which the left side refers to a classical metric tensor $g_{ab}$ and the right to the expectation value of the (renormalized) stress-energy tensor of quantum matter that propagates on the classical geometry defined by $g_{ab}$. It is often assumed that this theory would serve as an excellent approximation to full quantum gravity away from the Planck regime. In 3+1 dimensions, exact solutions to semi-classical gravity are difficult to come by and, more importantly, it has not been possible to check if they are good approximations because we do not have full quantum gravity. Let us examine this issue through the lens of our model where we can work out predictions of full quantum gravity and compare them with the semi-classical theory. A solution to semi-classical gravity consists in specifying a classical metric $g_{ab}$ and a quantum state of matter $|\Psi \rangle$ such that the quantum field $\hat{\Phi}$ satisfies its field equation with respect to $g_{ab}$ and the pair $(g_{ab},\,\, |\Psi\rangle)$ satisfies the semi-classical Einstein's equation. Choose a classical solution $(g_{ab}, \Phi)$ in our model. Then it follows that in the asymptotic region the pair $(g_{ab},\,\, \kcoh)$ solves semi-classical equations exactly. Thus, now we have an infinite class of solutions to semi-classical gravity. However, as we just saw, there are regimes --i.e. choices of $\phio(\o)$-- for which $g_{ab}$ is a poor approximation to full quantum geometry. And this occurs even in the asymptotic region where curvature vanishes identically! This suggests that the domain of validity of the semi-classical approximation can be quite subtle; it may not be dictated just by space-time curvature.

Interaction between quantum gravity and quantum information communities has increased over the last 5 years or so, giving rise to a healthy exchange of ideas. As mentioned in section \ref{s1}, this dialog has sparked interest in manifestation of quantum gravity effects that do not involve the `radiative degrees of freedom' of gravity but are instead induced by the quantum nature of matter sources (see, e.g. \cite{boseetal,cmvv,cbp,dgw}). Thus, they involve the `Coulombic' part of the gravitational field.  
As we discussed in section \ref{s2},\, 2+1 gravity is especially well suited to analyze these quantum effects because the gravitational field does not carry any radiative degrees of freedom. It is determined entirely by matter: while there is an infinite number of degrees of freedom, in our analysis they all reside in the matter sector. Therefore \emph{all} quantum properties of geometry/gravity are induced by the quantum properties of matter. We found that the metric has a number of quintessentially quantum properties. If the quantum state of matter $\kcoh$ is sharply peaked on a classical field $\mathring\Phi(R,T)$, we are in a regime that is tame from the perspective of quantum matter. Yet, the expectation value of $\hat{g}_{RR}$ can differ significantly from that of the classical solution determined by $\mathring\Phi (R,T)$. The difference is entirely due to non-perturbative quantum effects, and of course goes to zero in the $\hbar \to 0$ limit. There is also a trans-Planckian regime in which matter continues to behaves in a tame manner --its quantum state $\kcoh$ is sharply peaked at a classical field $\mathring\Phi(R,T) \sim \phio(\o)$ with very small fluctuations-- but the quantum metric has experiences very large quantum fluctuations. In both these regimes, one can carry out thought experiments. One can use spinning test particles in the asymptotic region where $\Phi(R,T)$ vanishes: there is no direct interaction between quantum matter that sources the gravitational field and test particles. One can parallel propagate the spinning particles around a closed loop $C$ and measure the change in its spin. This results in a measurement of the holonomy $U^{(C)}$ of the gravitational spin connection. Although the curvature of the physical metric vanishes everywhere along $C$, the holonomy is non-trivial --this is the gravitational analog of the well-known Aharonov-Bohm effect in electro-magnetism. Measurement of this holonomy determines the non-trivial metric component $g_{RR}= - g_{TT}$ as well as the Hamiltonian/energy of the \emph{total} system, matter+gravity. One can carry out repeated measurements and determine the expectation values and quantum uncertainties in these observables. Since these experiments are carried out entirely in the gravitational sector, they provide unambiguous test of the quantum nature of geometry/gravity, that is induced entirely by matter. As we saw, the specific non-linear coupling between matter and geometry of general relativity can enhance the small quantum fluctuations in the matter sector to produce large quantum effects in geometry. These effects are \emph{non-perturbative} and not considered in current discussions, where gravity is treated perturbatively and, furthermore, matter is generally modeled by non-relativistic particles, rather than quantum fields. Our framework is free of both these limitations. It provides a proof of principle that quantum information can be transferred to the gravitational sector from the matter sector through Coulombic interaction without any reference to gravitons or radiative modes. 

It is natural to ask if the effects that were uncovered by our analysis are specific to gravity or if they would also arise in electromagnetism.%
\footnote{I thank Norbert Straumann for raising this question. The calculation I summarize below resulted from that inquiry.} 
Are there interesting quantum effects on the non-radiative, Coulombic aspects of the Maxwell field by charged 
quantum fields that sources them? Now, as we saw, in 2+1 dimensions, Maxwell fields are dual to scalar fields and even with the axis-symmetric restriction, they carry radiative degrees of freedom. Therefore, in the 2+1 dimensional electromagnetic analog of our model the Maxwell field would not be determined entirely by charged sources; it would have its own radiative degree of freedom. A better candidate is provided by the spherically symmetric sector of 3+1 dimensional system consisting of Maxwell and charged Klein Gordon fields. In this sector, there are no photons, just as in our 2+1 dimensional model there are no gravitons. Therefore, if we use a charged Klein-Gordon quantum field $\hat\Phi$ as the source, the quantum Maxwell field $\hat{F}_{ab}$ is represented as an operator on the Hilbert space of $\hat\Phi$. Thus, any quantum features we find in the Maxwell field $\hat{F}_{ab}$ are induced by the quantum Klein-Gordon field via Coulomb interaction. It turns out that this model is also exactly soluble in Minkowski space-time. Spherical symmetry singles out a time-translation Killing field $t^a$ which one can use to decompose $F_{ab}$ into its electric and magnetic parts and introduce a canonical notion of frequency for the Klein-Gordon field. In the asymptotic region, we can express the only non-trivial component ${E}^r$ of the electric field in terms of the total charge ${Q}$ of the Klein-Gordon field, just as we could express $g_{RR} = - g_{TT}$ in terms of the total (Minkowski) energy $\Ho$ of matter sources in the gravitational case. We can again construct coherent states $\kcoh$ in the charged Klein-Gordon sector where (for simplicity) $\mathring\Phi$ has compact support on every Cauchy slice and calculate expectation values and fluctuations of $\hat{E}^r$ in the asymptotic region. They are non-zero. However, whereas the sophisticated non-linearities of Einstein dynamics in the 2+1 theory imply the relation $g_{RR} = e^{G\Ho}$, the 3+1 Maxwell theory implies that the relation between $E^r$ and $Q$ is linear, $E^r = \f{Q}{r^2}$. Therefore, while quantum fluctuations in the Klein-Gordon sector do induce quantum effects in the Maxwell sector even in absence of photons, they are not exponentially magnified as in the gravitational case. Indeed, the expressions one obtains in the Maxwell case resemble those in the weak field approximation of our 2+1 gravitational analysis.%
\footnote{In general relativity, the gravitational interaction `dresses' the bare mass to yield surprising results also in 3+1 dimensions. These are genuinely non-perturbative effects that are lost if one expands the result in powers of Newton's constant. For a discussion of this phenomenon, see, e.g., Chapter 1 of \cite{aa-book}.}

These concrete insights on a number of conceptual issues could be arrived at because the model is exactly soluble. However, the very reasons that make it exactly soluble also limit its reach. As explained in section \ref{s2}, from a 4-d perspective, the model represents Einstein-Rosen waves which have cylindrical symmetry. These symmetries make it totally unsuitable for the laboratory experiments that are being considered. Nonetheless, the analysis brought out a number of unforeseen effects and conceptual subtleties in 3-d quantum gravity coupled to matter. To what extent can one take them over to 4 dimensions? An obvious strategy would be to consider general relativity coupled to scalar fields in 4-d, but restrict oneself to the spherically symmetric sector. Then again, we would have an infinite dimensional midi-superspace where there are again no `radiative degrees of freedom' in the gravitational field. The metric would be determined entirely by matter and one can investigate quantum gravity effects that are induced by quantum matter, just as in the present case. But now, the analysis becomes significantly more complicated mathematically because one cannot decouple the scalar field dynamics from that of the metric.  
%
%\footnote{One can envisage changing the 4-d theory --as in the Callan-Giddings-Harvey-Strominger \cite{cghs} model-- in which one can again solve for first the scalar field on a fictitious flat space-time. But even in this simplified model, there is gravitational collapse leading to a black hole.}
%
From a physical perspective, there is now the possibility of a gravitational collapse leading to a black hole. While this is a rich sector, it has its own deep puzzles associated with the Hawking evaporation, and therefore not directly useful extract clean insights of the type that our 2+1 dimensional model provided. However, one can focus just on `sub-critical' initial data sets so that the scalar waves come in from infinity and scatter off to infinity. This restriction would prevent us from addressing interesting issues such as the Bekenstein bound \cite{jb} and related conjectures \cite{ls,gth} that led to the idea of holography. But one could still probe the status of other issues discussed in this section. That analysis will have to take in to account another key difference between the 2+1 and 3+1 dimensional general relativity. In 2+1 dimensions, the metric is flat outside sources while in the 3+1 context it only approaches the flat metric as $1/r$ as one recedes from the matter sources. Therefore, in 4-dimensional asymptotically flat situations, effects we found  --such as those due to the presence of trans-Planckian frequencies--  will decay as one moves away from sources and the fluctuations will be non-negligible only near the sources. However, this limitation is present also in the experiments that are currently contemplated. Thus, while the most striking predictions of the 3-d model will not carry over to 4-d situations, the subtleties brought out by the analysis may provide useful guidance. At a conceptual level, its detailed analysis unearthed directions in which one can look for novel effects, and at a mathematical level it suggests strategies to represent matter sources by quantum fields, rather than non-relativistic particles, and go beyond perturbation theory in the gravitational sector.

\section*{\bf Acknowledgements:} 
This work was supported by  the NSF grant PHY-1806356 and the Eberly Chair funds of Penn State. I would like to thank Fernando Barbero, Chris Beetle, Alex Corichi, Rodolfo Gambini, Badri Krishnan, Guillermo Mena, Monica Pierri, Jorge Pullin, Norbert Straumann, Thomas Thiemann and Madhavan Varadarajan for numerous discussions. Many of the results were presented at the Erwin Schr\"odinger Institute, Vienna; Schmidt-fest at the Albert Einstein Institute, Potsdam; Gravitational Waves Workshop at IIT Gandhinagar; and seminars elsewhere. I would like to thank the participants for their comments and questions that added clarity to the presentation of this work.

\end{document}